# Observation of self-organized criticality (SOC) behavior during edge biasing experiment on TEXTOR

Y. H. Xu, S. Jachmich, R. R. Weynants

*Laboratory for Plasma Physics, Ecole Royale Militaire/Koninklijke Militaire School, Euratom-Belgian State Association, Avenue de la Renaissance 30, B-1000 Brussels, Belgium, Partner in the Trilateral Euregio Cluster*

The self-organized criticality (SOC) behavior of the edge plasma transport has been investigated using the fluctuation data measured in the plasma edge and the scrape-off layer of TEXTOR tokamak before and during the edge electrode biasing experiments. In the "nonshear" discharge phase before biasing, both the potential and density fluctuations clearly exhibit some of the characteristics associated with SOC: (1) existence of $f^{-1}$ power-law dependence in the frequency spectrum, (2) slowly decaying long tails in the autocorrelation function, (3) values of Hurst parameters larger than 0.5 at all the detected radial locations, (4) non-Gaussian PDF of fluctuations and (5) radial propagation of avalanche-like events in the edge plasma area. During the biasing phase, with the generation of an edge radial electric field $E_r$ and hence a sheared $E_r$xB flow, the local turbulence is found to be well decorrelated by the $E_r$xB velocity shear, consistent with theoretical predictions. Nevertheless, it is concomitantly found that the Hurst parameters are substantially enhanced in the negative flow shear region and in the scrape-off layer as well, which is contrary to theoretical expectation. Implication of these observations to our understanding of plasma transport mechanisms is discussed.

**I. Introduction**

To understand some characteristics of plasma transport in magnetically confined plasmas, such as the large-size Bohm scaling behavior [1] departed from the diffusive model dominated by gyro-Bohm scaling [2], the self-organized criticality (SOC) hypothesis [3] has been proposed recently suggesting the existence of avalanche-type transport [4]. The central idea of SOC is long-range spatial and temporal correlation through scale invariance, by which avalanches can occur via nonlinear dynamical interaction [4-6]. In general, mixed diffusive and SOC dynamics will control transport [7]. In recent years, experimental evidence considered as key ingredients of SOC, such as long-range time correlations (or self-similarity) [8], empirical similarity of frequency spectra [9], intermittent behavior [10], radial propagation of avalanche-like events[11] and the self-similar distortion of Poisson-distributed quiet-times to the power law form [12] of plasma fluctuations has been widely found in a number of devices, although absence of long-range correlation was also detected elsewhere [13]. In this paper, we show the results of SOC behavior investigated on TEXTOR tokamak in the edge electrode biasing experiment [14-15]. Two goals were aimed at. First, we look for possible evidence of SOC behavior in the ohmically heated plasmas before biasing. This phase is slightly different from the "standard" ohmic discharge phase on TEXTOR. Because of the insertion of the electrode at zero voltage, the $E_r$ profile in the plasma boundary is flattened with roughly zero $E_r \times B$ shear rate, while in the "standard" ohmic discharge a naturally occurring $E_r \times B$ flow shear layer often exists around the last closed flux surface (LCFS). We will further on refer to this phase as the "non-shear" phase. Second, we explore how long and short time transport events are affected by a controlled sheared edge $E_r \times B$ flow during the biasing or "shear" phase. For this, we have analyzed the power spectra [$S(f)$], autocorrelation functions (ACF) and the Hurst parameters via the rescaled range (R/S) [16, 17] and structure function (SF) [18, 19] methods for the fluctuation data detected at different radial locations.

**II. Experimental conditions**

The experiments were carried out on TEXTOR under the following discharge conditions: R=175cm, $a \cong 48$cm, $B_T$=2.33T, $I_p$=200kA, $V_l$=1V and $\bar{n}_e = 1.0 \times 10^{19}$m$^{-3}$ in ohmically heated plasmas. The floating potential fluctuations were measured in both the plasma edge and the scrape-off layer (SOL) by a set of Langmuir probes consisting of carbon tips with 3.5mm in diameter. The probe pins are 3.75mm separated radially. The whole probe system can be moved radially from shot to shot. The fluctuation data were digitized at a rate of 500 kHz. For creating an edge electric field $E_r$ and hence a sheared $E_r \times B$ flow, a biasing voltage quickly ramped from 0 to 600~700V was applied between an inserted electrode located at r$\cong$41cm and the toroidal belt limiter (ALT-II) during the flat top of the discharge. Details on the experiments have been described in ref. [20]. Fig. 1 plots the radial dependences of the floating potential $\phi_f$ and the radial electric field $E_r$ detected by the probes before (black symbols and lines) and during (red symbols and lines) the biasing phase. From the figure, we can see that in the "non-shear" discharge phase before biasing, the $\phi_f$ and $E_r$ profiles in the plasma boundary are both flat with roughly zero $E_r$xB shear rate. During the biasing or "shear" phase, the $\phi_f$ profile is highly enhanced inside the LCFS, resulting in a narrow positive $E_r$ structure between r≈46.7cm and r≈48.3cm, as seen in Fig. 1(b).

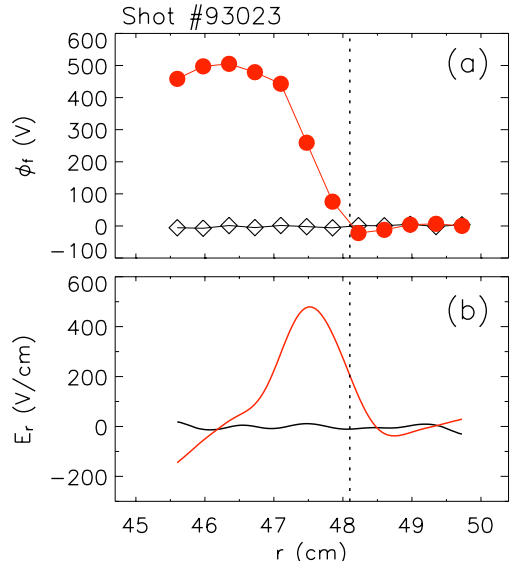

*Fig.1. Radial profiles of (a) the floating potential $\phi_f$ and (b) the radial electric field $E_r$ before (black color) and during (red color) the biasing discharge phase. The vertical dotted line marks the position of the last closed flux surface.*

**III. Results and discussion**

As mentioned above, for the present study of SOC dynamics, we have mainly analyzed the S(*f*), ACFs and Hurst parameters(*H*) via R/S and SF methods of the fluctuation data measured at different radial locations. The definitions of these notions and their links to the SOC dynamics have been given in details in ref. [20]. For this analysis, the dataset of each discharge was broken into eight unoverlapping subblocks of 8,000 points (total time lag =16ms). The S(*f*), ACF, R/S (as defined in Eq. (2) of ref. [20]) and SF (the average of the 1/*q* power of $S_{w,q}(\tau)$, $\langle S_{w,q}^{1/q} \rangle$, as defined in Eq. (5) of ref. [20]), for each subblock was calculated for the floating potential fluctuation data and then averaged over the eight subblocks. The statistical error of *H* is about 0.03. The stationarity test of the fluctuation data, using SF technique, indicates that the data are stationary within certain ranges and thus the *H* values can be properly determined in these ranges [19, 20]. The results measured at a location close to the LCFS in the "non-shear" phase are shown in Fig. 2. In Fig. 2(a), it is seen that the frequency spectrum, S(*f*), displays roughly three distinct regions in the range of *f*=(0.2-250) kHz with approximate decay indices of 0, -1 and –2, respectively. This result is in good agreement with other experimental observations [9, 21, 22] and resembles those obtained in the sandpile modeling and in turbulence model realizations of SOC systems [5, 23], i. e., (i) a high frequency part (scaled as $f^{-n}$, where *n* is 2 or higher) signifying small scale events involving very small parts of the system, (ii) a low frequency part (scaled nearly as $f^0$) reflecting single events with a global scale and (iii) the intermediate range (with $f^{-1}$ dependence), which has been related to the overlapping of avalanche transport. The corresponding ACF is plotted in Fig. 2(b). At small time lags the value of ACF drops very rapidly with time. The peak part of the ACF carries information on the correlation of local fluctuations. The *e*-folding time of the ACF, i. e., the width of the peak, $\tau_D$, is thus taken to be the decorrelation time of the local turbulence[24]. Fig. 2(b) shows that $\tau_D \approx 10\mu s$ for the present dataset. The existence of long-time correlation should be evident from an algebraic tail in the ACF [25]. Such a slow decay is clearly seen in Fig. 2(b), over large intervals of time lags until $\tau > 120\mu s$ ($\approx 12\tau_D$). For a quantitative determination of the self-similarity parameter, *H*, the R/S and SF methods are both applied for cross-checking the accuracy of the values obtained. The corresponding plot of R/S values (solid circles) versus time lags is shown in Fig. 2(c). From the figure, we can see that for time lags smaller than a few $\tau_D$ ($\tau < 100\mu s$), there is a transient process giving a slope of about 0.9. This non-stationary process identifies the local turbulence decorrelation behavior. At longer time scales, R/S settles on an "asymptotic" power law in the "self-similarity range", over which *H*=0.74 are determined from the slope (shown by dashed lines). Using SF, the slope of the SF curve in the stationary range can also give *H* directly. As seen in Fig. 2(c), the R/S and SF methods exhibit good agreement for getting *H*, consistent with simulations [19]. Similar features on S(*f*), ACF, R/S and SF have also been seen on the density fluctuation data [20], which further qualifies the universality of SOC paradigm. The radial profiles of the Hurst exponents estimated from potential fluctuations at various radial positions in the "non-shear" discharge phase are plotted in Fig. 5 (b) [see black circles]. It is noted that at all the detected radial locations, the Hurst parameters are well above 0.5, indicating the existence of long-range dependencies in the fluctuation dynamics in both the plasma edge and the SOL.

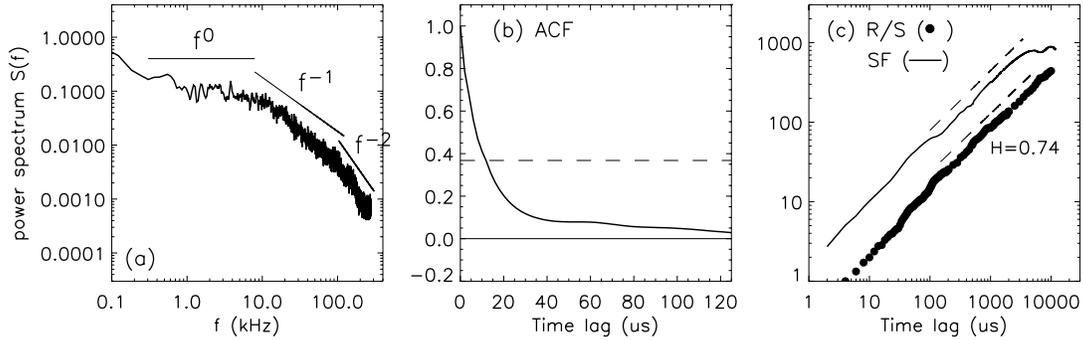

*Fig. 2. Power spectrum S(f), autocorrelation function (ACF) and R/S (SF) analysis of the floating potential fluctuations measured at r=48.1cm in the "non-shear" discharge phase. In (c), R/S values are shown by solid circles; the SF curve is the average of the 1/q power of $S_{w,q}(\tau)$ for q=0.5, 1, 2, 3, 4, 5; The dashed lines are the best fit to the R/S slope with H= 0.74; and the parallel ones show comparison with the SF curves.*

Further documentations of the turbulence in the edge of TEXTOR in the "non-shear" discharge phase, supporting the SOC hypothesis, are given in Figs. 3 and 4. Shown in Figs. 3(a) and (b) are the probability density function (PDF) of the floating potential and ion saturation fluctuations measured nearby the LCFS. For both signals, their PDFs representing positive fluctuations are clearly skewed, indicating fluctuations much greater than what is expected from a pure random distribution as reflected by the dotted Gaussian fit. This intermittent aspect on fluctuations, departing from the random diffusive model [2], explicitly suggests the existence of convective plasma transport. Similar results have also been obtained in other machines [10]. Moreover, evidence of large-scale spatially correlated events has been found in the plasma edge of TEXTOR. By using six radially spaced probes, the floating potential fluctuations in these pins are measured simultaneously. The cross-correlation function (CCF) of the fluctuations between the six probes is then calculated with a reference one at r=47.85cm. The contour plot of the pairwise CCF is shown in Fig. 4, where the motion of the maximum CCF to larger time lags reveals an outward propagation at an effective speed of $V_r \approx 330$m/s. This motion in the radial direction clearly identifies this feature as an avalanche. Similar phenomena are observed in DIII-D on temperature fluctuations [11]. Note that the radial extent of the coherent feature is ~22mm, much larger than the local gyroradius ($\rho_i \approx 0.4$mm), which is the typical scale length of gyro-Bohm scaling [2]. All these facts are consistent with plasma transport characterized by SOC dynamics.

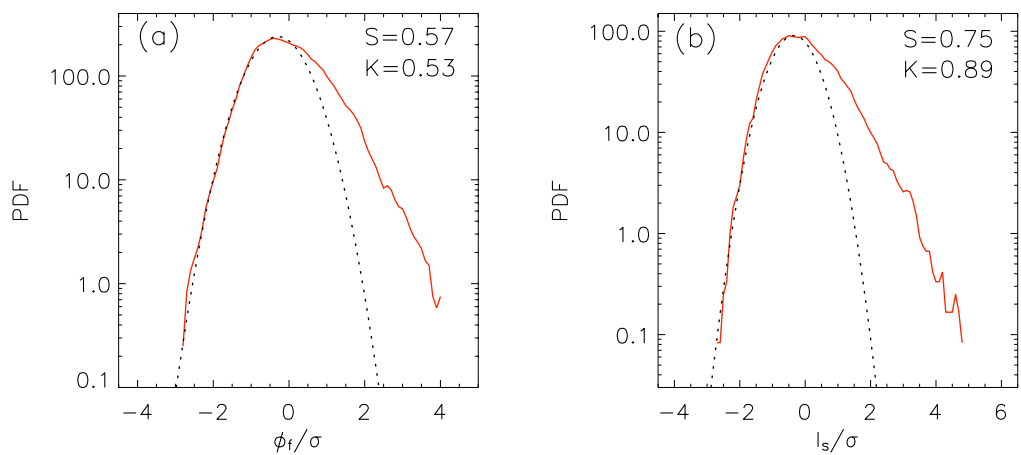

*Fig. 3. The probability density function (PDF) of (a) the floating potential fluctuations measured at r/a=1.02; and (b) ion saturation fluctuations measured at r/a=0.98. The dotted line in each plot is the best fit by a Gaussian function.*

To explore the shear-flow influence on SOC-like transport events, the long-term and short-term correlations are studied in the biasing experiment. Up to now, the predictions of shear flow effects on long-range time correlations are mainly based on the sandpile modeling, in which the R/S analysis indicates a decorrelation of avalanches when a sheared wind flow is included into the sandpile [26]. In W7-AS, evidence has been shown that the *H* values near the edge shear flow layer are slightly lower than those on either side [8], in agreement with the modeling. In our experiment, we have measured both the values of *H* and $\tau_D$ before and during the biasing phases. In Fig. 5, the radial dependencies of the $E_r$ shear, *H* and $\tau_D$ are plotted with black symbols denoting before and red symbols during the biasing phase, respectively. From Figs. 5(a) and (b), it can be seen that during the biasing phase the Hurst parameters are increased substantially in the SOL and in the negative $E_r$ shear zone as well. Meanwhile, we can see that the local decorrelation time, $\tau_D$, drops sharply in the negative and slightly in the positive $E_r$ shear region in the biasing phase (see Fig. 5(c)), in agreement with theoretical predictions [27]. The subsistence of long-range correlations in the flow shear region is surprising, and in contrast with the sandpile simulations [26]. Several factors might be thought of as possibly able to contaminate the H values. (i) Low frequency perturbations to increase the *H* values. In our case, no such perturbations either clearly show up in time traces of the raw signals (except for 25 Hz ringing of power supply, which has been filtered on the fluctuation data analyzed) nor in the stationarity tests of the fluctuation data during the biasing phase. (ii) The accurateness of Doppler-shift correction to affect *H* values. Such corrections may affect *H* in the high-flow zone (46.7cm<r<48.3cm). But in the outer SOL region (r>49cm), the flow is very small or close to zero and this correction cannot therefore play a role. One plausible explanation for the deviation with respect to the sandpile simulation may come from the modeling done in ref. 7, where the interplay between the diffusive and SOC transport dynamics is studied. For the dominant discontinuous SOC transport channel and the subdominant continuous diffusive channel, the modeling indicates that with the increase of diffusivity ratio, the 1/f region of the power spectrum shrinks and the Hurst exponent decreases as the continuous smoothing of the local inhomogeneities in the slope profile by the increased diffusion makes avalanches more difficult to take place. In contrast, at low ratio of diffusivity, the diffusive component cannot well balance the source at the submarginal level, causing the slope to build up so that large SOC-type avalanches can occur. This seems to be consistent with our experiment. In the "shear" phase, the local turbulence is strongly decorrelated, as $\tau_D$ is reduced dramatically in the negative $E_r$ shear region, and thus the SOC channel prevails. The present results demonstrate that an $E_r \times B$ flow shear alone, at least in our case, is not sufficient to suppress avalanche-like transport. The question as to the conditions necessary to effectively decorrelate SOC-type avalanches by sheared flow appears therefore still unanswered. The fact that the Hurst parameters are almost unchanged during the biasing phase in the positive $E_r$ shear locations, as seen in Fig. 5(b), remains also to be explained.

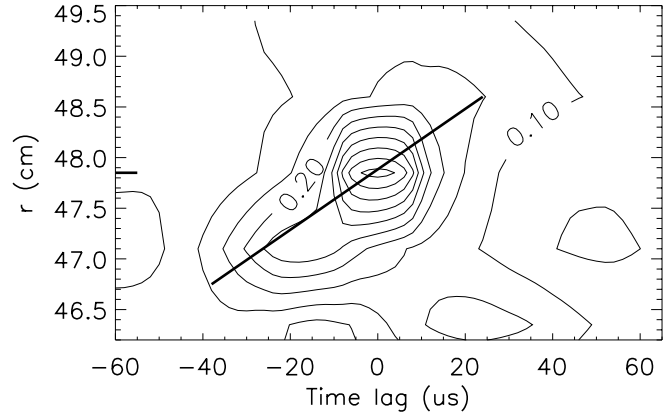

*Fig. 4. Contours of the pairwise cross-correlation function (CCF) measured at plasma edge in the "non-shear" discharge phase before biasing. The motion of maximum CCF indicates a radially outward propagation in the plasma edge.*

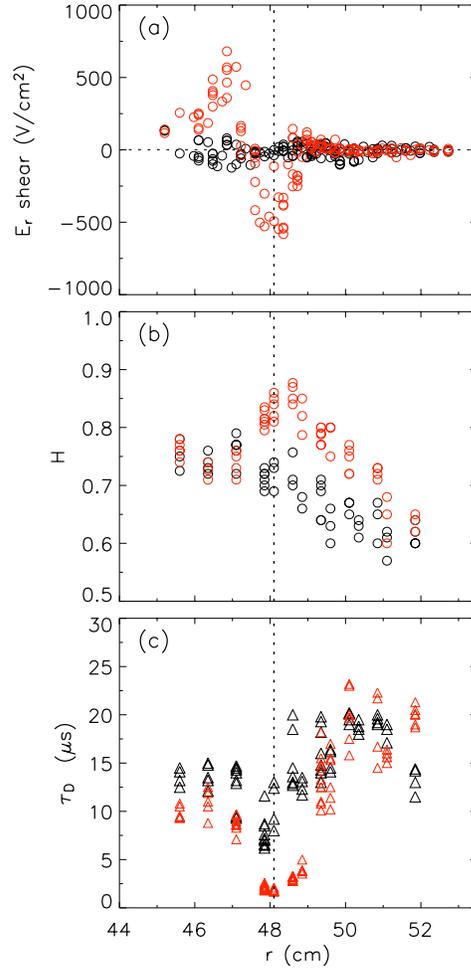

Fig. 5. Radial profiles of (a) radial electric field $E_r$ shear, (b) Hurst parameter (H) estimated by R/S, and (c) local decorrelation time of turbulence ($\tau_D$), before (black symbols) and during (red symbols) the biasing phase. The vertical dotted line marks the position of the last closed flux surface.

**IV. Conclusion**

In conclusion, the fluctuation data measured at the plasma edge and the SOL of TEXTOR in the edge biasing experiments have been analyzed for the study of SOC-relevant phenomena. Evidence has been found that, in the "non-shear" discharge phase before biasing, the frequency spectrum exhibits $f^{-1}$ power-law dependence; the autocorrelation function displays long decaying tail; the R/S and SF analyses show self-similarity parameters well larger than 0.5 at all measured locations; the PDF shows a non-Gaussian distribution; and CCF identifies a radial propagation of avalanche-like events in the edge plasma area. All these results support the idea that SOC dynamics may play a significant role in plasma transport. During the biasing phase of the edge electrode polarization experiment, with the generation of an edge $E_r$x$B$ velocity shear, we found that the Hurst exponents are enhanced substantially in the negative $E_r$ shear zone and also in the SOL. The results indicate that an $E_r$x$B$ flow shear alone, at least in our case, is not sufficient to suppress the SOC-type transport. $E_r$x$B$ velocity shear, especially in the negative $E_r$ shear region, can however very well decorrelate the local turbulence. The diffusive transport channel is thereby strongly suppressed. We tend to conclude that (i) in the ohmic phase, SOC and diffusive transport channels coexist in the TEXTOR edge; while, with strong shear, SOC transport prevails, and (ii) in the negative $E_r$ shear region, "*the dynamics governing the decorrelation of the local fluctuations and the long-range time dependencies are probably different, one being the decorrelation of the turbulence and the other being the decorrelation of the transport events (avalanches)*" as stated in Ref. 8.

**References:**

[1] ITER Physics Basis on "Plasma confinement and transport", Nucl. Fusion **39**, 2175(1999).

[2] W. Horton, Rev. Mod. Phys. **71**, 735 (1999); B. B. Kadomtsev, *Plasma Turbulence* (Academic, London, 1965).

[3] P. Bak, C. Tang, and K. Wiesenfeld, Phys. Rev. Lett. **59**, 381 (1987).

[4] P. H. Diamond and T. S. Hahm, Phys. Plasmas **2**, 3640 (1995).

[5] D. E. Newman, B. A. Carreras, P. H. Diamond, and T. S. Hahm, Phys. Plasmas **3**, 1858 (1996).

[6] X. Garbet and R. E. Waltz, Phys. Plasmas **5**, 2836 (1998).

[7] R. Sánchez, D. E. Newman, B. A. Carreras, Nucl. Fusion **41,** 247 (2001).

[8] B. A. Carreras, B. Ph. van Milligen, M. A. Pedrosa *et al.,* Phys. Rev. Lett. **80**, 4438 (1998).

[9] M. A. Pedrosa, C. Hidalgo, B. A. Carreras *et al.*, Phys. Rev. Lett. **82**, 3621 (1999).

[10] G. Y. Antar, S. I. Krasheninnikov, P. Devynck *et al.*, Phys. Rev. Lett. **87**, 065001(2001).

[11] P. A. Politzer, Phys. Rev. Lett. **84**, 1192 (2000).

[12] R. Sánchez, B. Ph. van Milligen, D. E. Newman and B. A. Carreras, Phys. Rev. Lett. **90**, 185005 (2003).

[13] B. A. Carreras, B. Ph. van Milligen, M. A. Pedrosa *et al.,* Phys. Plasmas **5**, 3632 (1998).

[14] R. R. Weynants, G. V. Oost, G. Bertschinger *et al.*, Nucl. Fusion **32**, 837 (1992).

[15] S. Jachmich, G. V. Oost, R. R. Weynants *et al.*, Plasmas Phys. Control. Fusion **40**, 1105 (1998).

[16] H. Hurst, Trans. Am. Soc. Civ. Eng. **116**, 770 (1951).

[17] B. B. Mandelbrot and J. R. Wallis, Water Resour. Res. **5**, 967 (1969).

[18] A. Davis, A. Marshak, W. Wiscombe, and R. Cahalan, J. Geophys. Res. **99**, 8055 (1994).

[19] C. X. Yu, M. Gilmore, W. A. Peebles, and T. L. Rhodes, Phys. Plasmas **10**, 2772 (2003).

[20] Y. H. Xu, S. Jachmich, R. R. Weynants *et al.*, to be published in Phys. Plasmas **11**(12), 2004.

[21] T. L. Rhodes, R. A. Moyer, R. Groebner *et al.*, Phys. Lett. A **253**, 181 (1999).

[22] B. A. Carreras, R. Balbin, B. Ph. van Milligen *et al.,* Phys. Plasmas **6**, 4615 (1999).

[23] T. Hwa and M. Kadar, Phys. Rev. A **45**, 7002 (1992).

[24] C. P. Ritz, H. Lin, T. L. Rhodes, and A. J. Wootton, Phys. Rev. Lett. **65**, 2543 (1990).

[25] J. Beran, *Statistics for Long-Memory Processes* (Chapman and Hall, New York, 1994).

[26] D. E. Newman, B. A. Carreras, and P. H. Diamond, Phys. Lett. A **218**, 58 (1996).

[27] H. Biglari, P. H. Diamond and P. W. Terry, Phys. Fluids **B2**, 1 (1990).